\documentclass[preprint,tightenlines,aps,prd,groupedaddress,showpacs]
              {revtex4}%
\usepackage{graphicx}
\usepackage{amsmath}
\usepackage{bm}
\usepackage{relsize}
\RequirePackage{xspace}
\begin{document}

\title{
Color-evaporation-model calculation of \\
$\bm{e^+ e^-\to J/\psi+c}\bar{\bm{c}}\bm{+X}$ at $\bm{\sqrt{s}=10.6}$~GeV}

\author{Daekyoung Kang, Jong-Wan Lee, and Jungil Lee}
\affiliation{
Department of Physics, Korea University, Seoul 136-701, Korea}

\author{Taewon Kim }
\affiliation{
Department of Physics, KAIST, Daejon 305-701, Korea}

\author{Pyungwon Ko}
\affiliation{
School of Physics, KIAS, Seoul 130-722, Korea}



\date{\today}
\begin{abstract}
Measurements by the Belle Collaboration of the
cross section for inclusive $J/\psi$ production in $e^+e^-$ annihilation
have been a serious challenge to current heavy-quarkonium theory.
Especially, the measured cross sections for exclusive $J/\psi+\eta_c$ and
inclusive $J/\psi+c\bar{c}+X$ differ from nonrelativistic QCD predictions 
by an order of magnitude. In order to check if other available alternative 
theory
can resolve such a large discrepancy, we calculate the cross section for
inclusive $J/\psi+c\bar{c}+X$ based on the color-evaporation model. 
As a phenomenological model, the color-evaporation model is still 
employed to predict cross sections for inclusive quarkonium production
in various processes.
Our results show that the color-evaporation-model prediction is even 
smaller than the nonrelativistic QCD prediction by an order of magnitude.
The resultant color-evaporation-model prediction 
for $J/\psi+c\bar{c}+X$ fraction in
the inclusive $J/\psi$ production cross section is 0.049, while 
the empirical value measured by the Belle Collaboration is 0.82.
\end{abstract}


\maketitle

Spin-triplet $S$-wave charmonium states such as $J/\psi$ have been clean probes
of both perturbative and nonperturbative feature of quantum 
chromodynamics~(QCD).
Systematic studies based on first principles became possible after the 
introduction of the nonrelativistic QCD(NRQCD) factorization 
formalism~\cite{BBL}, an effective field theory of QCD. However, still
there are several challenges to quarkonium physics research. 
One of the most interesting problems is regarding inclusive 
$J/\psi$ production in $e^+e^-$ annihilation at $B$ factories.
The Belle Collaboration has measured the cross section for exclusive
$J/\psi + \eta_c$ production by observing a peak in the momentum spectrum of 
inclusive $J/\psi$ signals that corresponds to the 2-body final state 
$J/\psi + \eta_c$~\cite{Abe:2002rb}. The measured cross section is by an 
order of magnitude larger than the predictions of 
NRQCD~\cite{Braaten:2002fi,Liu:2002wq}. Interesting 
proposals~\cite{Bodwin:2002fk,Bodwin:2002kk,Brodsky:2003hv} to resolve 
the problem were disfavored by a recent analysis by the Belle 
Collaboration~\cite{Abe:2004ww}. Having failed in explaining the 
result using lowest-order perturbative calculations,
one may guess that higher-order corrections in strong coupling
$\alpha_s$ may be very large~\cite{Hagiwara:2003cw}. If it is true,
perturbative expansion is not a proper method to predict the cross section.
Recently, we proposed that the measurement of the cross section for 
the $e^+ e^-$ annihilation into four 
charm hadrons may provide us with a strong constraint in determining
the origin of the large discrepancy~\cite{Kang:2004mz} and corresponding
experimental analysis is being carried out.
In the Belle analysis~\cite{Abe:2002rb} they also reported the cross section 
for the inclusive $J/\psi+c\bar{c}$ production at the center-of-momentum(c.m.) 
energy $\sqrt{s}=10.6~$GeV. 
From the measured inclusive cross sections
$\sigma(e^+e^-\to J/\psi +D^{*+}+X)=0.53\,^{+0.19}_{-0.15}\pm 0.14$~{pb}
and
$\sigma(e^+e^-\to J/\psi +D^{0}+X)=0.87\,^{+0.32}_{-0.28}\pm 0.20$~{pb}
they extracted the cross section 
for $J/\psi+c\bar{c}+X$~\cite{Abe:2002rb}
based on the Lund model for fragmentation of a charm quark into the $D$ 
mesons~\cite{Sjostrand:1993yb}.
The resulting cross section is
$\sigma(e^+e^-\to J/\psi +c\bar{c}+X)%
=0.87\,^{+0.21}_{-0.19}\pm 0.17$~{pb}~\cite{Abe:2002rb},
which is larger than NRQCD 
predictions~\cite{Cho:1996cg,Yuan:1996ep,Baek:1998yf} 
by an order of magnitude. For example, an NRQCD prediction 
given in Ref.~\cite{Cho:1996cg} is about 0.07~pb.
The Belle Collaboration also calculated the ratio 
$\sigma(e^+e^-\to J/\psi +c\bar{c}+X)/\sigma(e^+e^-\to J/\psi +X)=%
0.82\pm 0.15\pm 0.14$~\cite{Abe:2002rb,Update}. 
This remarkable result reveals the fact that $e^+e^-\to J/\psi +c\bar{c}$ 
is the dominant source for inclusive $J/\psi$ production~\cite{Abe:2001za} in 
$e^+e^-$ annihilation at $B$ factories.  Again, the result is remarkably
larger than available NRQCD predictions. 
Two-photon mediated process
$e^+e^-\to J/\psi +c\bar{c}$ of order $\alpha^4$~\cite{Liu:2003zr}
has been calculated using NRQCD in order to
check if the photon-fragmentation effect is significant like
the process $e^+e^-\to J/\psi+J/\psi$~\cite{Bodwin:2002fk,Bodwin:2002kk}.
Unlike the exclusive double $J/\psi$ production
the photon-fragmentation contribution was found to be negligible 
compared to the single-photon process of order $\alpha^2\alpha^2_s$
in $e^+e^-\to J/\psi +c\bar{c}$~\cite{Liu:2003zr}.
Different approaches were also introduced using a large 
$K$-factor~\cite{Hagiwara:2004pf}
and the nonperturbative quark-gluon-string model~\cite{Kaidalov:2003wp}. 

In this paper, we calculate the cross section for
inclusive $e^+e^-\to J/\psi+c\bar{c}+X$ process at the c.m. 
energy $\sqrt{s}=10.6~$GeV using
the color-evaporation model(CEM)~\cite{Fritzsch:1977ay,%
Halzen:1977rs,Gluck:1977zm,Barger:1979js}.
Predictions of the quark-hadron-duality model, which is similar to the
CEM, are available~\cite{Kiselev:1994pu,Berezhnoy:2003hz}
and they underestimate the cross section for $e^+e^-\to J/\psi+c\bar{c}$ 
like that of NRQCD.
The quark-hadron-duality model assumes that only a color-singlet $c\bar{c}$ 
state evolves into a $J/\psi$, while the CEM sums over all possible spin
and color states. Thus the CEM calculation will probe the importance of
the missing color-octet component. 
The hard process for the inclusive production of $J/\psi+c\bar{c}$ is 
approximated by $e^+e^-\to c\bar{c}c\bar{c}$ in leading order in strong
coupling $\alpha_s$.  Invariant mass of the $c\bar{c}$ pair evolving into 
the final $J/\psi$ is restricted from $2m_c$ to open charm threshold $2m_D$. 
Multiplying the CEM parameter $F_{J/\psi}$ to the hard-scattering cross section,
the CEM prediction for the  inclusive $J/\psi+c\bar{c}$ production cross 
section is obtained. The CEM parameter $F_{J/\psi}$ represents the probability
of the $c\bar{c}$ pair evolving into a $J/\psi$.
Our results show that the CEM prediction is significantly smaller than
that of NRQCD, which already underestimates the empirical cross section
by an order of magnitude. The process is well distinguished from many other
hadronic processes, where CEM predictions are not that far away from NRQCD ones.
Comprehensive reviews on the CEM can be found in 
Refs.~\cite{Brambilla:2004wf,Bedjidian:2003gd}.

In the CEM the cross section for inclusive $J/\psi$ production is
obtained by integrating differential cross section for
$c\bar{c}+X$ over the invariant mass $m_{c\bar{c}}$
of the $c\bar{c}$ pair from $2m_c$ to $D\bar{D}$ threshold, $2m_D$.
\begin{equation}
\label{eq:sig-psi}
\sigma(J/\psi+X)=
F_{J/\psi} \int_{2m_c}^{2m_D}
\left(\frac{d\sigma_{c\bar{c}}}{dm_{c\bar{c}}}\right)dm_{c\bar{c}},
\end{equation}
where $\sigma_{c\bar{c}}$ is the cross section for
$e^+e^-\to c\bar{c}+X$.
The CEM parameter $F_{J/\psi}$ in Eq.~(\ref{eq:sig-psi}) represents 
the probability for the $c\bar{c}$ pair to evolve into the final 
$J/\psi$ inclusively.
This  phenomenological parameter is fitted to the data from hadronic 
experiments which have a large number of data samples.  
Thus the uncertainties of the parameter 
come from parton distribution function and renormalization/factorization
scale as well as the charm-quark mass $m_c$. 

Before we proceed to consider the inclusive $J/\psi+c\bar{c}+X$ production, 
we must address the point that the CEM is a model valid only for inclusive
production of a heavy quarkonium. The parton cross section $\sigma_{c\bar{c}}$
in Eq.~(\ref{eq:sig-psi}) includes all possible spin and color states for a
$c\bar{c}$ pair. The final quarkonium state is always accompanied by light hadrons
because the quantum numbers for $c\bar{c}$ and the quarkonium can be different in 
general. Therefore, it is impossible to apply the CEM to predict the exclusive 
two-charmonium production in $e^+e^-$ annihilation.

\begin{figure*}
\includegraphics[width=11.5cm]{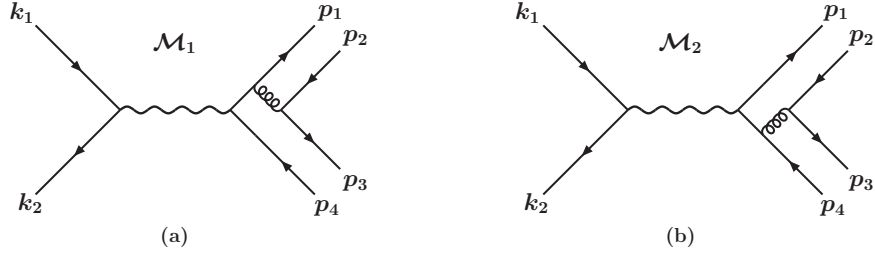}
\caption{\label{fig1}
Two topologically distinct
Feynman diagrams for
$e^-(k_1) e^+ (k_2)\to c(p_1) \bar{c}(p_2) c(p_3) \bar{c}(p_4)$
among 8 diagrams. The pair $c(p_1) \bar{c}(p_2)$ is to evolve into
the final $J/\psi$.
}
\end{figure*}

In order to apply the CEM to calculate the cross section for the inclusive 
$J/\psi+c\bar{c}$ production in $e^+e^-$ annihilation, we have to calculate the
invariant-mass distribution of a $c\bar{c}$ pair created in the process
$e^+e^-\to c\bar{c}c\bar{c}+X$ and follow the way shown in 
Eq.~(\ref{eq:sig-psi}).
In leading order in strong coupling $\alpha_s$, 
$c\bar{c}c\bar{c}$ can be produced at order $\alpha^2\alpha_s^2$.
There are two topologically distinct Feynman diagrams generating 
two pairs of
$c\bar{c}$, which are shown as $\mathcal{M}_1$ and $\mathcal{M}_2$
in Fig.~\ref{fig1}(a) and \ref{fig1}(b), respectively.
Momenta for the involving particles are assigned as
$e^-(k_1) e^+ (k_2)\to c(p_1) \bar{c}(p_2) c(p_3) \bar{c}(p_4)$.
The amplitude for the two diagrams shown in Fig.~\ref{fig1} are
\begin{equation}
-i\mathcal{M}_i=
i\frac{(4\pi)^2 e_c\alpha \alpha_s}{s(p_2+p_3)^2}
\,\bar{v}_e(k_2)\gamma_\alpha u_e(k_1)
\,\bar{u}(p_3)T^a\gamma_\beta v(p_2)
\,\bar{u}(p_1)T^a H_i^{\alpha\beta}v(p_4),
\label{eq:amp}
\end{equation}
where $s=(k_1+k_2)^2$,
$e_c=\frac{2}{3}$ is the fractional electric charge of the
charm quark, and $a$ is the SU(3) color index for the virtual gluon.
The vector indices $\alpha$ and $\beta$ are for the virtual photon
and the gluon, respectively. 
We suppress the spin and color indices of 
the charm quarks in Eq.~(\ref{eq:amp}).
For $i=1$ or 2 the tensors ${H}_i^{\alpha\beta}$ in Eq.~(\ref{eq:amp}),
which are matrices in spinor space, are defined by
\begin{subequations}
\begin{eqnarray}
H_1^{\alpha\beta}&=&\gamma^\beta\Lambda(p_1+p_2+p_3)\gamma^\alpha,
\\
H_2^{\alpha\beta}&=&\gamma^\alpha\Lambda(-p_2-p_3-p_4)\gamma^\beta,
\end{eqnarray}
\end{subequations}
where $\Lambda(p)=(\,/\!\!\!p+m_c)/(p^2-m_c^2)$.
There are 6 more Feynman diagrams that can be obtained from
the two amplitudes $\mathcal{M}_1$ and $\mathcal{M}_2$
by exchanging two charm quarks and two antiquarks, respectively, as
\begin{equation}
\begin{array}{ll}
 \mathcal{M}_3=-P_{1\leftrightarrow 3}\mathcal{M}_1,
&\mathcal{M}_4=-P_{1\leftrightarrow 3}\mathcal{M}_2,
\\
 \mathcal{M}_5=-P_{2\leftrightarrow 4}\mathcal{M}_1,
&\mathcal{M}_6=-P_{2\leftrightarrow 4}\mathcal{M}_2,
\\
 \mathcal{M}_7=+P_{1\leftrightarrow 3}P_{2\leftrightarrow 4} \mathcal{M}_1,
&\mathcal{M}_8=+P_{1\leftrightarrow 3}P_{2\leftrightarrow 4} \mathcal{M}_2.
\end{array}
\label{eq:amp3-8}
\end{equation}
where $P_{i\leftrightarrow j}$ is the operator exchanging two particles
with momentum indices $p_i$ and $p_j$ shown in Fig.~\ref{fig1}.
The signs of $\mathcal{M}_3$ through $\mathcal{M}_8$ in
Eq.~(\ref{eq:amp3-8}) are determined by the antisymmetry 
of Fermi statistics in exchanging identical fermions among 
the final-state particles.

In the process $e^+e^-\to c\bar{c}c\bar{c}$,
there are four ways to pick a pair of $c\bar{c}$, which will evolve
into the final $J/\psi$ inclusively. We choose 
the $c\bar{c}$ pair with momenta $p_1$ and $p_2$. 
Then we integrate over the invariant mass 
$m_{12}=\sqrt{(p_1+p_2)^2}$ of the pair from $2m_c$ to $2m_D$. 
The invariant mass $m_{34}=\sqrt{(p_3+p_4)^2}$ 
of the other pair is integrated 
over the whole phase space, which runs from $2m_c$ to $\sqrt{s}-2m_{12}$.
Because we distinguish two charm quarks and two anti-charm quarks
simultaneously, we do not multiply the statistical factor $(1/2)^2$ 
for identical particles in the phase space integration. 
Corresponding CEM prediction for the inclusive process
$e^+e^-\to c\bar{c}c\bar{c}+X$ is calculated using the formula
\begin{equation}
\sigma(J/\psi+c\bar{c}+X)=
F_{J/\psi} \int_{2m_c}^{2m_D}
\left(\frac{d\sigma}{dm_{12}}\right)
dm_{12},
\label{eq:sig-psicc}
\end{equation}
where the differential cross section 
$\frac{d\sigma}{dm_{12}}$ in Eq.~(\ref{eq:sig-psicc}) is defined by
\begin{equation}
\frac{d\sigma}{dm_{12}}=
\frac{1}{32(2\pi)^8 s^{3/2}}
\int_{2m_c}^{\sqrt{s}-m_{12}} dm_{34}\int  d\Omega\,
 d\Omega^*_{12}\,
 d\Omega^*_{34}\,
|\bm{P}| |\bm{p}_1^*| |\bm{p}_3^*|
\overline{\sum}\left|\mathcal{M}\right|^2,
\label{eq:dsig}
\end{equation}
where $\mathcal{M}=\sum_{i=1}^8 \mathcal{M}_i$.
The summation notation $\overline{\sum}$ in Eq.~(\ref{eq:dsig}) 
stands for 
averaging over initial spin states and summation over 
final color and spin states.
The solid angle $d\Omega$ is for the three momentum $\bm{P}$ of
$p_1+p_2$ in the $e^+e^-$ c.m. frame.
Remaining two solid-angle elements of the form $d\Omega^*_{ij}$ 
are for the three momentum $\bm{p}^*_i$ in the rest frame of $p_i+p_j$,
where $\{i,j\}=\{1,2\}$ or $\{3,4\}$. 
Integrating over the eight variables in Eqs.~(\ref{eq:sig-psicc}) 
and (\ref{eq:dsig}), 
we get the CEM prediction for 
the cross section $\sigma(e^+e^-\to J/\psi+c\bar{c}+X)$.
The cross section for $e^+e^-\to c\bar{c}c\bar{c}$ can be obtained
if we remove the $F_{J/\psi}$ in Eq.~(\ref{eq:sig-psicc}),  
replace the cut $2m_D$ in Eq.~(\ref{eq:sig-psicc}) by $\sqrt{s}-2m_c$, 
and multiply the statistical factor $(1/2)^2$ to consider two pairs 
of identical particles in the final state. 
The process $e^+e^-\to c\bar{c}c\bar{c}$ has been considered
in Ref.~\cite{Kang:2004mz}.
We compute the 
$\overline{\sum}|\mathcal{M}|^2$ in Eq.~(\ref{eq:dsig})
using REDUCE~\cite{reduce} 
and carry out the phase-space integral in Eqs.~(\ref{eq:sig-psicc})
and (\ref{eq:dsig})
making use of the adaptive Monte Carlo routine VEGAS~\cite{vegas}.
As a check, we carry out the same calculation using 
CompHEP~\cite{Boos:2004kh}.
Our analytic result for $\overline{\sum}|\mathcal{M}|^2$ and
numerical values for the total cross section agree with
those obtained by using CompHEP.

Let us summarize our theoretical inputs.
Following recent quarkonium calculations
in Refs.~\cite{Braaten:2002fi,Bodwin:2002fk,Bodwin:2002kk},
we use the electromagnetic coupling $\alpha=1/137$ and 
the next-to-leading order pole mass $m_c=1.4\pm 0.2$~GeV for the charm-quark mass. 
The strong coupling constant $\alpha_s$ depends on the typical scale in the 
hard-scattering cross section. If we use the renormalization scale 
$\mu=2m_c$ with $m_c=1.4$~GeV, the value is $\alpha_s(2m_c)\approx 0.26$. 
The value is appropriate for the gluon-fragmenation contribution.
However, a $c\bar{c}$ pair from different quark lines can evolve into the $J/\psi$. 
In this case,  the momentum transfer carried by the internal gluon can be as 
large as $\sqrt{s}/2=5.3$~GeV to get $\alpha_s(\sqrt{s}/2)\approx 0.21$. 
The latter value has been used for the exclusive $J/\psi+\eta_c$ production in 
Ref.~\cite{Braaten:2002fi} because the momentum scale is exactly $\sqrt{s}/2$.
In our numerical analysis we take into account the uncertainty in the scale 
and allow the variation $\alpha_s=0.23\pm 0.03$.
Because the cross section is proportional to $\alpha_s^2$, the uncertainty
from the strong coupling is about $\pm 30$\%.
Finally, we have to fix the numerical value
for the CEM parameter $F_{J/\psi}$ for directly produced $J/\psi$. 
The CEM parameter $F^{\textrm{inc.}}_{J/\psi}$ for inclusive $J/\psi$ production,
which includes feed-down from higher charmonium resonances, has been fitted to 
the data from $pp$ and $pA$ collisions. They have small statistical errors
but depend on patron distribution functions, renormalization/factorization 
scale, and $m_c$. A recent set of the values can be found in
Refs.~\cite{Vogt:2002vx,Vogt:2002ve,Bedjidian:2003gd,Brambilla:2004wf}.
In Table 3.7 of Ref.~\cite{Bedjidian:2003gd} several values are given depending
on parton distribution function and $m_c$. The values are
$F^{\textrm{inc.}}_{J/\psi}=0.0248$ for $m_c=1.4$~GeV, 
$0.0229$ for $m_c=1.3$~GeV, and
$0.0144 \sim 0.0155$ for $m_c=1.2$~GeV. The fraction of direct $J/\psi$ is 0.62
as shown in Table 3.6 of Ref.~\cite{Bedjidian:2003gd}.
The values for directly produced $J/\psi$ are then
$F_{J/\psi}=0.015$ for $m_c=1.4$~GeV, $0.014$ for 
$m_c=1.3$~GeV, and about $0.01$ for $m_c=1.2$~GeV.
In Ref.~\cite{Amundson:1995em}, authors extracted the parameter using
the open charm production in the CEM framework to estimate the QCD correction
to the hadroproduction of charmonium.
The values in  Ref.~\cite{Amundson:1995em} are consistent with the photoproduction
data~\cite{Amundson:1995em,Amundson:1996qr}.
For $m_c=1.45$~GeV, the authors of
Ref.~\cite{Amundson:1995em,Amundson:1996qr} found $F^{\textrm{inc.}}_{J/\psi}=(0.43\sim 0.5)/9$ 
which results in $F_{J/\psi}=0.030\sim 0.034$.
Roughly taking into account the difference between the two fits, we take
$F_{J/\psi}= 0.025 \pm 0.010$ for $m_c=1.4$~GeV. 
The uncertainty from the CEM parameter is about $\pm 40$\%. 
The fits in Refs.~\cite{Amundson:1995em,Amundson:1996qr} favor larger values, 
while the fits in  Ref.~\cite{Bedjidian:2003gd} prefer smaller ones.
As shown in Ref.~\cite{Bedjidian:2003gd} numerical value for $F_{J/\psi}$ 
depends on $m_c$. We take into account the $m_c$ dependence of the 
$F_{J/\psi}$ described above by making a linear fit to the values given 
in the Table 3.7 of Ref.~\cite{Bedjidian:2003gd}. 
Resultant fit is 
$F_{J/\psi}=0.025+0.047\left(\frac{m_c}{\textrm{GeV}}-1.4\right)$.

\begin{figure}
\includegraphics[width=8cm]{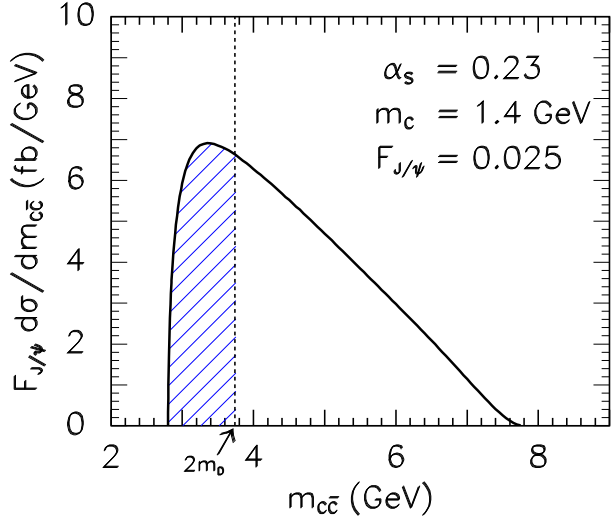}
\caption{\label{fig2}
Differential cross section
$F_{J/\psi}d\sigma/dm_{c\bar{c}}$ in fb/GeV
with respect to the invariant mass $m_{c\bar{c}}=m_{12}$
of $c\bar{c}$ for $e^+ e^-$ annihilation into $c\bar{c}c\bar{c}$,
where $m_c=1.4$~GeV,  $\alpha=1/137$, $\alpha_s=0.23$, and
$F_{J/\psi}=0.025$.
The area of the shadowed region with $2m_c\le m_{c\bar{c}}\le 2m_D$
stands for the CEM prediction for $\sigma(e^+e^-\to J/\psi+c\bar{c}+X)$.
}
\end{figure}

\begin{figure}
\includegraphics[width=8cm]{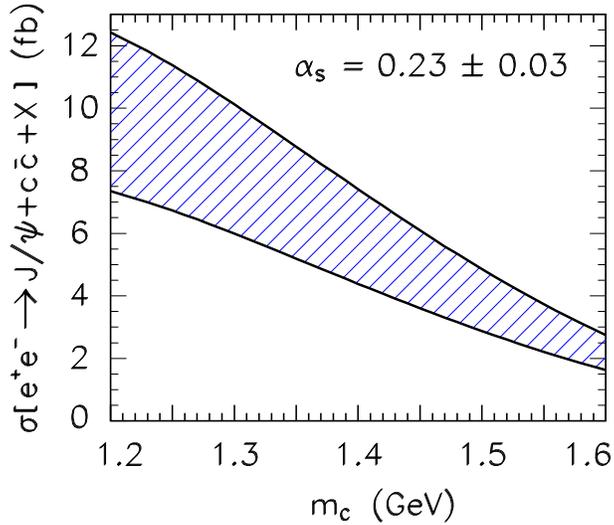}
\caption{\label{fig3}
The CEM prediction for the inclusive cross section
$\sigma(e^+e^-\to J/\psi+c\bar{c}+X)$
in fb at $\sqrt{s}=10.6$~GeV as a function of $m_c$.
The $m_c$-dependent CEM parameter 
$F_{J/\psi}=0.025+0.047\left(\frac{m_c}{\textrm{GeV}}-1.4\right)$
is used and the band represents the uncertainty from the strong coupling
 $\alpha_s=0.23\pm0.03$ only.
}
\end{figure}

In Fig.~\ref{fig2} we show the differential cross section
$d\sigma/dm_{12}$ in Eq.~(\ref{eq:dsig}) multiplied by the CEM parameter 
$F_{J/\psi}=0.025$ with  $m_c=1.4$~GeV and $\alpha_s=0.23$.
The area of the shadowed region in Fig.~\ref{fig2} corresponds to the 
inclusive cross section $\sigma(e^+e^-\to J/\psi+c\bar{c}+X)$. 
If we increase the $m_c$, the cross section decreases because the left 
end point of the phase space for $m_{12}$ shifts toward the cut 
$2m_D=2\times 1.87$~GeV and the height of the peak decreases.
The cross section gets larger if we choose a smaller $m_c$.
The $m_c$ dependence of our CEM predictions for the inclusive cross section 
$\sigma(e^+e^-\to J/\psi+c\bar{c}+X)$ at $\sqrt{s}=10.6~$GeV
is shown in Fig.~\ref{fig3}. The band represents the uncertainty 
$\approx \pm 30$\% coming from the strong coupling 
$\alpha_s=0.23\pm 0.03$. In Fig.~\ref{fig3} we use the 
$m_c$-dependent CEM parameter 
$F_{J/\psi}=0.025+0.047\left(\frac{m_c}{\textrm{GeV}}-1.4\right).$
Considering roughly $\pm 40$\% uncertainty in $F_{J/\psi}$ and
$\pm 30$\% uncertainty in strong-coupling dependence, our prediction has
roughly $\pm 50$\% uncertainty neglecting $m_c$ dependence, which is well
described in Fig.~\ref{fig3}.

In previous NRQCD calculations for $e^+e^-\to J/\psi+c\bar{c}+X$ 
authors used larger values for the strong coupling 
$\alpha_s$~\cite{Cho:1996cg,Yuan:1996ep,Baek:1998yf}. 
For example, $\alpha_s(2m_c)=0.28$ was used in Ref.~\cite{Cho:1996cg}
to get the NRQCD prediction for the cross section 
$\sigma(e^+e^-\to J/\psi+c\bar{c}+X)=0.07~$pb at
$\sqrt{s}=10.6~$GeV. In order to
have a fair comparison with the NRQCD predictions let us replace
our $\alpha_s$ with $\alpha_s=0.28$. 
Then our predictions given in Figs.~\ref{fig2} and \ref{fig3}
should be increased by a factor of $(0.28/0.23)^2\approx 1.5$. 
The upper bound of our CEM cross section is about $12$~fb
which is less than the NRQCD prediction by a factor of $6$.
The underestimation of the cross section in the CEM may be due to the limited
phase space for the invariant mass $m_{c\bar{c}}$ because of the small
center-of-momentum energy. Further studies of invariant-mass spectra 
for hadronic processes will reveal if our guess is valid.

Another important feature of our result is that the CEM prediction is
less than that of the quark-hadron-duality model. The prediction of
the quark-hadron-duality model is about 
60~fb~\cite{Kiselev:1994pu,Berezhnoy:2003hz} which is similar to 
the NRQCD predictions~\cite{Cho:1996cg,Yuan:1996ep,Baek:1998yf}. 
The quark-hadron-duality model collects only 
color-singlet contributions while the CEM includes all the color degrees 
of freedom. In the CEM the invariant mass of the $c\bar{c}$ pair is
integrated over $2m_c<m_{c\bar{c}}<2m_D$, which is less than the 
region $2m_c<m_{c\bar{c}}<2m_{D}+(0.5\sim 1~\textrm{GeV})$ for
the quark-hadron-duality model.
If we increase the upper limit for the $m_{c\bar{c}}$ by 1~GeV,
the CEM cross section at $m_c=1.4$~GeV becomes 12~fb, 
which is still smaller than the quark-hadron-duality-model prediction 
by a factor of 5. 
A reason for this difference may be due to the long-distance factor 
$F_{J/\psi}$. We use the value for the  $F_{J/\psi}$ extracted from  
hadronic collisions, while the quark-hadron-duality model uses a different 
method.

Before closing our discussion, let us comment on the ratio 
$\sigma(e^+e^-\to J/\psi +c\bar{c}+X)/\sigma(e^+e^-\to J/\psi +X)=%
0.82\pm 0.15\pm 0.14$ measured by the Belle Collaboration~\cite{Update}.
The empirical value is remarkably larger than the prediction of NRQCD.
Unlike NRQCD, the nonperturbative factor $F_{J/\psi}$ in the CEM prediction 
for the ratio cancels and the ratio is depending only on $\alpha_s$ and $m_c$.
Therefore CEM prediction for the ratio should be more reliable than
the absolute value for the cross sections.
The inclusive $J/\psi$ production rate can be calculated by considering
the parton process $e^+e^-\to c\bar{c}g$ in the leading order in $\alpha_s$.
Note that $e^+e^-\to c\bar{c}$ channel is forbidden because 
$m_{c\bar{c}}=\sqrt{s}\gg 2m_D$.  According to our recent analysis~\cite{new}, 
CEM prediction for the cross section is
\begin{equation}
\sigma(e^+e^-\to J/\psi +X)\approx 95~\textrm{fb},
\end{equation}
where we use $F_{J/\psi}=0.025$ and $\alpha_s(2m_c)=0.26$ with $m_c=1.4~$GeV. 
Resulting ratio becomes
\begin{equation}
\left.
\frac{\sigma(e^+e^-\to J/\psi+c\bar{c} +X)}
     {\sigma(e^+e^-\to J/\psi +X)}
\right|_{\rm CEM}
\approx0.062.
\label{R:CEM}
\end{equation}
The empirical rate measured by the Belle Collaboration was obtained 
with the cut $|\mathbf{p}^*_{J/\psi}|>2.0$~GeV 
of the three-momentum of the $J/\psi$
in the $e^+e^-$ c.m. frame. If we impose the same cut to our prediction,
our prediction for the ratio becomes $\approx$ $0.049$, which is less than 
the value given in Eq.~(\ref{R:CEM}) by about 21\%.
The ratio is again significantly smaller than the experimental
value $\approx 0.82$.

In summary, we have calculated the CEM prediction for the inclusive cross 
section for $e^+ e^-\to J/\psi+c\bar{c}+X$
using the CEM parameter $F_{J/\psi}$ fitted to the data from $pp$ and $pA$ 
collisions~\cite{Vogt:2002vx,Vogt:2002ve,Bedjidian:2003gd,Brambilla:2004wf}
and from photoproduction~\cite{Amundson:1996qr}.
Our CEM prediction for the cross section 
$\sigma(e^+ e^-\to J/\psi+c\bar{c}+X)$ is about 
$9.7$~fb at $m_c=1.2$~GeV, $5.8$~fb at $m_c=1.4$~GeV, and 
$2.2$~fb at $m_c=1.6$~GeV.
The cross section is, at least by a factor of 6, smaller than
the NRQCD prediction, which is already smaller than the
empirical value by an order of magnitude.
Thus the color-evaporation model severely underestimates
the inclusive $J/\psi+c\bar{c}+X$ cross section 
in $e^+e^-$ annihilation measured by the Belle Collaboration.

\begin{acknowledgments}
We would like to thank Geoff Bodwin for introducing us the problem
studied in this work. We also thank Eric Braaten and Geoff Bodwin 
for valuable suggestions.
PK is supported in part by 
KOSEF through CHEP at Kyungpook National University.
JL is supported by a Korea Research Foundation Grant(KRF-2004-015-C00092).
\end{acknowledgments}



\begin{thebibliography}{}
\bibitem{BBL}
G.~T.~Bodwin, E.~Braaten, and G.~P.~Lepage,
	Phys.\ Rev.\ D {\bf 51}, 1125 (1995);
{\bf 55}, 5853(E) (1997).


\bibitem{Abe:2002rb}
K.~Abe {\it et al.}  [BELLE Collaboration],
Phys.\ Rev.\ Lett.\  {\bf 89}, 142001 (2002).

\bibitem{Braaten:2002fi}
E.~Braaten and J.~Lee,
Phys.\ Rev.\ D {\bf 67}, 054007 (2003)
[arXiv:hep-ph/0211085].

\bibitem{Liu:2002wq}
K.~Y.~Liu, Z.~G.~He and K.~T.~Chao,
Phys.\ Lett.\ B {\bf 557}, 45 (2003)
[arXiv:hep-ph/0211181].

\bibitem{Bodwin:2002fk}
G.~T.~Bodwin, J.~Lee and E.~Braaten,
Phys.\ Rev.\ Lett.\  {\bf 90}, 162001 (2003)
[arXiv:hep-ph/0212181].

\bibitem{Bodwin:2002kk}
G.~T.~Bodwin, J.~Lee, and E.~Braaten,
Phys.\ Rev.\ D {\bf 67}, 054023 (2003)
[arXiv:hep-ph/0212352].

\bibitem{Brodsky:2003hv}
S.~J.~Brodsky, A.~S.~Goldhaber and J.~Lee,
Phys.\ Rev.\ Lett.\  {\bf 91}, 112001 (2003)
[arXiv:hep-ph/0305269].

\bibitem{Abe:2004ww}
K.~Abe {\it et al.}  [Belle Collaboration],
Phys.\ Rev.\ D {\bf 70}, 071102 (2004)
[arXiv:hep-ex/0407009].


\bibitem{Hagiwara:2003cw}
K.~Hagiwara, E.~Kou and C.~F.~Qiao,
Phys.\ Lett.\ B {\bf 570}, 39 (2003)
[arXiv:hep-ph/0305102].

\bibitem{Kang:2004mz}
D.~Kang, J.-W.~Lee, J.~Lee, T.~Kim, and P.~Ko,
arXiv:hep-ph/0412224, Phys. Rev. D, in press.

\bibitem{Sjostrand:1993yb}
T.~Sjostrand,
Comput.\ Phys.\ Commun.\  {\bf 82}, 74 (1994).

\bibitem{Cho:1996cg}
P.~L.~Cho and A.~K.~Leibovich,
Phys.\ Rev.\ D {\bf 54}, 6690 (1996)
[arXiv:hep-ph/9606229].
\bibitem{Yuan:1996ep}
F.~Yuan, C.~F.~Qiao and K.~T.~Chao,
Phys.\ Rev.\ D {\bf 56}, 321 (1997)
[arXiv:hep-ph/9703438].
\bibitem{Baek:1998yf}
S.~Baek, P.~Ko, J.~Lee and H.~S.~Song,
J.\ Korean Phys.\ Soc.\  {\bf 33}, 97 (1998)
[arXiv:hep-ph/9804455].
\bibitem{Update}K. Abe \textit{et. al} [Belle Collaboration],
BELLE-CONF-0331, contributed paper, International Europhysics Conference
on High Energy Physics (EPS2003), Aachen, Germany (2003).

\bibitem{Abe:2001za}
K.~Abe {\it et al.}  [BELLE Collaboration],
Phys.\ Rev.\ Lett.\  {\bf 88}, 052001 (2002).
\bibitem{Liu:2003zr}
K.~Y.~Liu, Z.~G.~He and K.~T.~Chao,
Phys.\ Rev.\ D {\bf 68}, 031501 (2003)
[arXiv:hep-ph/0305084].

\bibitem{Hagiwara:2004pf}
K.~Hagiwara, E.~Kou, Z.~H.~Lin, C.~F.~Qiao and G.~H.~Zhu,
Phys.\ Rev.\ D {\bf 70}, 034013 (2004)
[arXiv:hep-ph/0401246].

\bibitem{Kaidalov:2003wp}
A.~B.~Kaidalov,
JETP Lett.\  {\bf 77}, 349 (2003)
[Pisma Zh.\ Eksp.\ Teor.\ Fiz.\  {\bf 77}, 417 (2003)]
[arXiv:hep-ph/0301246].
\bibitem{Fritzsch:1977ay}
H.~Fritzsch,
Phys.\ Lett.\ B {\bf 67}, 217 (1977).
\bibitem{Halzen:1977rs}
F.~Halzen,
Phys.\ Lett.\ B {\bf 69}, 105 (1977).
\bibitem{Gluck:1977zm}
M.~Gluck, J.~F.~Owens and E.~Reya,
Phys.\ Rev.\ D {\bf 17}, 2324 (1978).
\bibitem{Barger:1979js}
V.~D.~Barger, W.~Y.~Keung and R.~J.~N.~Phillips,
Phys.\ Lett.\ B {\bf 91}, 253 (1980).
\bibitem{Kiselev:1994pu}
V.~V.~Kiselev, A.~K.~Likhoded and M.~V.~Shevlyagin,
Phys.\ Lett.\ B {\bf 332}, 411 (1994)
[arXiv:hep-ph/9408407].

\bibitem{Berezhnoy:2003hz}
A.~V.~Berezhnoy and A.~K.~Likhoded,
Phys.\ Atom.\ Nucl.\  {\bf 67}, 757 (2004)
[Yad.\ Fiz.\  {\bf 67}, 778 (2004)]
[arXiv:hep-ph/0303145].



\bibitem{Brambilla:2004wf}
N.~Brambilla {\it et al.},
arXiv:hep-ph/0412158.
\bibitem{Bedjidian:2003gd}
M.~Bedjidian {\it et al.},
arXiv:hep-ph/0311048.

\bibitem{reduce}
A.~C.~Hearn, \textit{REDUCE User's Manual} v. 3.7 (The RAND Corporation,
Santa Monica, 1999) (Email:reduce@rand.org).
\bibitem{vegas}
G.~P.~Lepage, J. Comput. Phys. \textbf{27}, 192 (1978).
\bibitem{Boos:2004kh}
E.~Boos {\it et al.}  [CompHEP Collaboration],
Nucl.\ Instrum.\ Meth.\ A {\bf 534}, 250 (2004)
[arXiv:hep-ph/0403113].
\bibitem{Vogt:2002vx}
R.~Vogt,
arXiv:hep-ph/0203151.

\bibitem{Vogt:2002ve}
R.~Vogt,
Heavy Ion Phys.\  {\bf 18}, 11 (2003)
[arXiv:hep-ph/0205330].
\bibitem{Amundson:1995em}
  J.~F.~Amundson, O.~J.~P.~Eboli, E.~M.~Gregores and F.~Halzen,
  Phys.\ Lett.\ B {\bf 372}, 127 (1996)
  [arXiv:hep-ph/9512248].

\bibitem{Amundson:1996qr}
  J.~F.~Amundson, O.~J.~P.~Eboli, E.~M.~Gregores and F.~Halzen,
  Phys.\ Lett.\ B {\bf 390}, 323 (1997)
  [arXiv:hep-ph/9605295].


\bibitem{new}
D.~Kang, J.-W.~Lee, J.~Lee, and T.~Kim,
in preparation.
\end{thebibliography}
\end{document}